\documentclass[a4paper,12pt]{article}
\pdfoutput=1
\usepackage{cite}
\usepackage{amsmath}
\usepackage{amsfonts}
\usepackage{amssymb}
\usepackage{graphicx, rotating}
\usepackage{epsfig}
\usepackage{latexsym}
\usepackage{graphicx}
\usepackage{color}
\usepackage{amsmath,bm,amssymb}

\usepackage{color}
\usepackage[english]{babel}
\usepackage{hyperref}

\newcommand{\Slash}[1]{{\ooalign{\hfil#1\hfil\crcr\raise.167ex\hbox{/}}}}

\newcommand{\beq}{\begin{equation}}  \newcommand{\eeq}{\end{equation}}
\newcommand{\bef}{\begin{figure}}  \newcommand{\eef}{\end{figure}}
\newcommand{\bec}{\begin{center}}  \newcommand{\eec}{\end{center}}
  
\newcommand{\laq}[1]{\label{eq:#1}}  

\newcommand{\Eq}[1]{Eq.~(\ref{eq:#1})}

\newcommand{\eq}[1]{(\ref{eq:#1})}

\def\({\left(}
\def\){\right)}

\def\O{\mathcal{O}}

\newcommand{\AND}{~{\rm and}~}
\newcommand{\EV}{ {\rm \, eV} }
\newcommand{\KEV}{ {\rm \, keV} }

\newcommand{\GEV}{ {\rm \, GeV} }

\def\d{\delta}

\def\f{\phi}
\def\g{\gamma}

\def\l{\lambda}
\def\m{\mu}
\def\n{\nu}

\def\s{\sigma}
\def\t{\tau}

\def\D{\Delta}
\def\G{\Gamma}
\def\F{\Phi}

\def\tl{\tilde}
\def\*{\dagger}

\begin{document}
\renewcommand\bibname{\Large References}

\begin{center}
\begin{flushright}
TU-1166
\end{flushright}

\vspace{1.0cm}

{\Large\bf  Indirect Detection
of eV Dark Matter \\ 
via Infrared Spectroscopy}
\vspace{1.0cm}

{\bf  Taiki Bessho$^{1}$, Yuji Ikeda$^{2,1}$ and Wen Yin$^{3}$}

\vspace{12pt}
\vspace{0.5cm}
{\em 
$^{1}$PhotoCross Inc., 17-203 Iwakura-Minami, Sakyo-ku, Kyoto 606-003, Japan\\
$^{2}${Laboratory of Infrared High-resolution Spectroscopy, \\Koyama Astronomical Observatory, Kyoto Sangyo \,\\ Motoyama, Kamigamo, Kita-ku, Kyoto
603-8555, Japan} \\
$^{3}${Department of Physics, Tohoku University, Sendai, Miyagi 980-8578, Japan } 
}

\vspace{0.5cm}
\abstract{
Infrared spectroscopy has been developed significantly. In particular, infrared photons can be measured with high spectral and angular resolution in state-of-art spectrographs.  They are sensitive to monochromatic photons due to the decay and annihilation of particles beyond the Standard Model, such as dark matter (DM), while insensitive to background photons that form a continuous spectrum. In this paper, we study the indirect detection of the DM decaying into infrared light using infrared spectrographs.  In particular, we show that serious thermal and astrophysical noises can be overcome. As concrete examples, the Warm INfrared Echelle spectrograph to Realize Extreme Dispersion and sensitivity  (WINERED) installed at the Magellan Clay 6.5m telescope and Near-Infrared Spectrograph (NIRSpec) at the James Webb Space Telescope (JWST) are discussed. We show that a few hours of measurements of a faint dwarf spheroidal galaxy with WINERED (NIRSpec-like spectrograph) in the Magellan telescope (JWST) can probe an axion-like particle DM in the mass range  $m_\phi=1.8 - 2.7\,$eV ($0.5-4\,$eV) with a photon coupling $g_{\phi\gamma\gamma}\gtrsim 10^{-11}{\rm GeV}^{-1}$. Complemental approaches, taking advantage of the high resolutions, such as the measurement of the Doppler shift of the signal photon lines and the possible search of the DM decay around the Milky Way galaxy center with Infrared Camera and Spectrograph (IRCS) at 8.2m Subaru telescope, are also presented. 
  }

\end{center}
\clearpage

\setcounter{page}{1}
\setcounter{footnote}{0}

%\tableofcontents

%%%%%%%%%%%%%%%%%%%%%%%%%%%%%%%%%%%%%%
\section{Introduction}

The origin of dark matter (DM) is one of the leading mysteries of particle theory, cosmology, and astronomy. 
The DM is known to be a long-lived, neutral, and cold matter that composes a dominant fraction of the present Universe. 
However, the particle origin of the DM, such as the mass or its coupling to the visible particles, i.e., those living in the standard model (SM) of the particle theory, is not known. So far, many theoretical and experimental efforts have been made to explore this origin.

The DM mass may be small because it can easily explain the long lifetime by the suppressed decay rate. 
 This kind of DM is slowly decaying in the present Universe. For instance,
the QCD axion, which solves the strong CP problem~\cite{Peccei:1977hh,Peccei:1977ur,Weinberg:1977ma,Wilczek:1977pj}, in the hadronic axion window has a mass around $\EV$~\cite{Chang:1993gm, Moroi:1998qs}. Such a QCD axion can be produced non-thermally consistent with the cold DM paradigm with a low reheating temperature~\cite{Moroi:2020has, Moroi:2020bkq, Nakayama:2021avl}.
Alternatively, the hypothesis that the inflaton and DM are unified by a single axion-like particle (ALP), i.e. ALP=DM=inflaton hypothesis, predicts the ALP 
mass to be around $\O(0.01-1)$ eV with photon coupling $g_{\f\g\g}\sim 10^{-11}\GEV^{-1}$~\cite{Daido:2017wwb, Daido:2017tbr} (see also \cite{Armengaud:2019uso}). 
The eV DM with $g_{\f\g\g}\sim 10^{-11}\GEV^{-1}$ are also phenomenologically interesting. 
 It is hinted from the extra-galactic background light (EBL) by two independent analyses, via the TeV gamma-ray spectrum and the  EBL anisotropy~\cite{Gong:2015hke, Korochkin:2019qpe,  Caputo:2020msf} (see also \cite{Kohri:2017oqn, Kalashev:2018bra, Kashlinsky:2018mnu, Nakayama:2022jza}). In addition, the ALP produced in the Horizontal branch star may explain the cooling hint~\cite{Ayala:2014pea, Straniero:2015nvc, Giannotti:2015kwo} 
(Reviews for axions or ALPs are given in Refs.~\cite{Jaeckel:2010ni,Ringwald:2012hr,Arias:2012az,Graham:2015ouw,Marsh:2015xka,Irastorza:2018dyq,DiLuzio:2020wdo}).
It may be important to experimentally search for the DM in the eV mass range. 

Turning now to the experimental side, significant progress has been made in recent years in the field of infrared light measurement techniques, particularly infrared spectroscopy. 
For instance, the Near Infrared Camera and the Near-Infrared Spectrograph (NIRSpec) play important roles in obtaining the scientific results from the James Webb Space Telescope (JWST), which is deployed in a solar orbit near the Sun-Earth $L_2$ Lagrange point.\footnote{see  https://jwst.nasa.gov/index.html.} 
 In particular, the NIRSpec at the JWST can measure an infrared spectrum with a spectral resolution of $R=2700$ and an angular resolution of $0.01~\rm arcsec^{2}$ in the wavelength range of $0.7-5\m$m. 
 Here 
 \beq
R \equiv \frac{\l}{\D \l_{\rm FWHM}}
 \eeq
with $\D \l_{\rm FWHW}$ being the peak full width at half of the maximum height. 
Although the main target of the JWST is to measure the red-shifted light from $\O(10)$ billion light-years away, 
in this paper, we study the indirect detection of the DM around the nearby galaxies within a million light-years
 by employing the state-of-the-art spectrographs 
including the NIRSpec. 

 Another important state-of-the-art spectrograph is the Warm INfrared Echelle spectrograph to Realize Extreme Dispersion and sensitivity (WINERED)~\cite{ikeda2006winered, yasui2008warm, kondo2015warm, ikeda2016high, ikeda2018very, 2022WINERED}. This is a near-infrared high-resolution spectrograph developed by the University of Tokyo and the Laboratory of Infrared High-resolution Spectroscopy. 
WINERED is a PI-type instrument. It was used with the $3.58$m New Technology Telescope (NTT) at the La Silla Observatory from 2017, and installed in the $6.5$m  Magellan Clay telescope at the Las Campanas Observatory. 
WINERED provides the highest sensitivity (the instrumental throughput as high as 50\%) among various high-resolution spectrographs attachable to the intermediate (3-4~m class) or the large (8-10~m class) telescopes in the short NIR region (0.9-1.35$\m$ m) distinctive features. 
The angular resolution is $0.29~{\rm arcsec}^{2}$ when attached to the Magellan telescope.  
In addition, WINERED has three observing modes: ``WIDE" ($R=28000$), “HIRES-Y” ($R=68000$), and “HIRES-J” ($R=68000$).

In this paper, we study the indirect detection of decaying DM into two particles, including a photon, by using infrared spectrographs. In particular, we use the WINERED in the Magellan Telescope and NIRSpec at JWST as concrete examples. 
 It is the extremely good spectral and angular resolution that makes them powerful detectors of the localized line-like spectrum, which is the special feature of the photon from a 2-body-decay of the DM in a dwarf spheroidal galaxy (dSph).
The high resolution suppresses the background signal from continuous background light, like the zodiacal light and atmospherical thermal radiation, in each bin while does not suppress the signal from the monochromatic component from a localized object. 
For instance a 4 hour measurement of a proper dSph in the Magellan telescope (JWST) will set the $2\s$ exclusion bound for a $1.8-2.7\EV$ ($0.5-4\EV$) ALP DM, $g_{\f\g\g} \gtrsim 10^{-11}\GEV^{-1}$ with the WINERED (NIRSpec-like) spectrographs. 
The future reach of the ALP measurement of Segue 1, an ultra-faint dSph, is shown in Figs.\ref{fig:para} and \ref{fig:para2}.  As one can see, they can cover the hadronic QCD axion region, ALP=DM=inflaton region, as well as the horizontal branch star/EBL hint in the corresponding mass range, if the exposure time is reasonably long. 

The photons from the Milky Way galaxy center and galaxy clusters, on the other hand, are no more line-like for the ``detectors" due to the Doppler shift from the virialized velocities of the DM. Moreover, there will be interstellar absorption, suppressing the signal flux. 
We argue that the Infrared Camera and Spectrograph (IRCS) at the 8.2m Subaru telescope~\cite{tokunaga1998infrared,kobayashi2000ircs} may overcome those difficulties for the search of the DM around the Milky Way galaxy center.  
Another approach that takes advantage of the high spectral resolution may be measuring the Doppler shift of the line-like lights due to the known radial velocities of several dSphs.

Before moving to the main part, let us refer to some relevant works. 
The indirect detection search with heavier DM is one of the best approaches to the origin of the DM. 
The decay of the DM to X and $\g$-ray has been well studied~\cite{Gruber:1999yr,Bouchet:2008rp,kappadath1998measurement,Strong:2004de,Fermi-LAT:2012edv}. 
The authors of \cite{Grin:2006aw,Regis:2020fhw} have derived the indirect detection bound from optical spectrographs using real experimental data. The mass range and spectrographs we propose to use in this paper are complementary to the previous ones. In particular, we will show that the crucial noises of the infrared thermal radiation and the sky background that are not (needed to be) discussed in the previous studies can be overcome in various state-of-art infrared spectrographs in the already realized proper environments, making it possible to probe the eV DM. 

The red-shifted infrared photons from the decay of the DM in the early Universe have been constrained from the EBL and the cosmic microwave background distortion~\cite{Overduin:2004sz, Jaeckel:2010ni,Ringwald:2012hr}. 
The analysis of the angular power spectrum at high $\ell$ of the cumulative light was used to approach the decay of several eVs DM~\cite{Gong:2015hke, Kohri:2017oqn, Kalashev:2018bra, Korochkin:2019qpe,  Caputo:2020msf, Shirasaki:2021yrp, Nakayama:2022jza}, thanks to the good angular resolution in the various experiments~\cite{Matsumoto:2010pz, Windhorst:2010ib, Kashlinsky:2012zz,  Mitchell-Wynne:2015rha, Seo:2015fga, Helgason:2016xoc}. This makes it possible 
to suppress the galactic foregrounds at high $\ell$. In contrast to them, we propose to use infrared spectrographs that also have a good spectral resolution to identify the line-like photon signal from the decay of the DM in the present Universe against the suppressed background signals.

This paper is organized as follows. In the next section, we discuss the sensitivity reach of the WINERED and NIRSpec-like spectrographs to detect
 the eV ALP DM by looking at dSphs. In  Sec. \ref{sec:comp}, we discuss the complementary possibilities for the spectrographs to detect eV DM. The last section is devoted to conclusions and discussion.

\section{Looking for eV ALP DM in dSphs with infrared spectrograghs} 

\subsection{ALP DM decaying into a pair of photons}
Let us consider the ALP DM, $\f$. The ALP is a pseudo-real scalar that couples to a pair of photons via the Lagrangian
\beq
{\cal L}=- \frac{g_{\f \g\g} }{4 }\f  F_{\mu\n}\tl{F}^{\m\n}, 
\eeq
where $F_{\m\n}$  ($\tl{F}^{\m\n}$) is the field strength of the SM photon (its dual), and $g_{\f \g\g} $ is the photon coupling. 
Via the interaction, the non-relativistic ALP can decay into two photons, $\f\to \g\g,$ with photon energy $E_\g = m_\f/2$ at the moment of the decay. Here  $m_\f$ is the mass of the ALP. 
The decay rate can be estimated as 
\beq
\G_\f= \frac{g_{\f\g\g}^2}{64\pi} m_\f^3. 
\eeq
In particular, a special relationship is 
\beq
\laq{QCDaxion}
g_{\phi \g\g}\simeq \(1.6-400\)\times10^{-11}\GEV^{-1} \(\frac{m_\f}{2\EV}\) {~~[\text{QCD axion}]}
\eeq 
which is the prediction of the QCD axion~\cite{Peccei:1977hh,Peccei:1977ur,Weinberg:1977ma,Wilczek:1977pj, GrillidiCortona:2015jxo}. We take the lower limit of the axion-photon coupling from the hadronic axion window~\cite{Chang:1993gm, Moroi:1998qs}. This is the case the photon coupling is accidentally canceled with the model-dependent parameter $E/N=2$~\cite{Moroi:1998qs}.\footnote{More precisely, we take the lowest value to be the hadronic uncertainty when this accidental cancellation happens~\cite{GrillidiCortona:2015jxo}.}  We take $E/N=12$ for the upper bound, the choice for which is not relevant in this paper. 

Another special relation is from the hypothesis that the ALP plays both the roles of DM and inflaton in the cosmology~\cite{Daido:2017wwb, Daido:2017tbr} (see also \cite{IAXO:2019mpb}). The requirement of the successful inflation gives
\beq \laq{ALP}
g_{\phi \g\g}\simeq c_\g 10^{-11}\GEV^{-1}\(\frac{2\EV }{m_\f } \)^{1/2} [{\rm ALP=DM=inflaton}]
\eeq
with $c_\g$ being the model-dependent parameter. We consider the range $c_\g=0.1-10$ as the natural region. 
Further requiring the successful cosmology without further assumptions, the mass is predicted in the range $m_\f=\O(0.01-1)\EV$ from the constraint of small-scale structures, successful reheating, and the DM abundance production. The existence of the viable parameter region against the various constraints is non-trivial, and the minimal scenario is called the ALP miracle. In this paper, we consider wider physics targets with more generic cosmology for the ALP=DM=inflaton scenarios (c.f. Refs.\,\cite{Takahashi:2021tff,Takahashi:2020uio}), and thus we only focus on the relation \Eq{ALP}. 

We also comment that there is a model-independent constraint from the cooling of the horizontal branch stars \cite{Raffelt:1985nk,Raffelt:1987yu,Raffelt:1996wa, Ayala:2014pea, Straniero:2015nvc, Giannotti:2015kwo, Carenza:2020zil} $g_{\f\g\g}\lesssim 6.6\times 10^{-11}\GEV^{-1}.$ (See also a recent paper~\cite{Dolan:2022kul} for a severer bound.)
This is applied if $m_\f\lesssim 10\KEV$. For the eV ALP DM, this is one of the most stringent bounds, and thus it is important if an ALP DM search reaches beyond this bound.

The generic ALP DM model predicts that the ALP DM is decaying in the present Universe.   The resulting differential flux of photons from the decay is composed of two components,
\beq
\frac{d\F_\g}{d E_\g}=   \frac{d\F_\g^{\text{extra}}}{d E_\g}+\sum_i\frac{d\F_{\g,i}}{d E_\g}
\eeq
Here the first term represents the extragalactic component, which is isotropic and the spectrum has a scaling of $\frac{d\F_\g^{\text{extra}}}{d E_\g}\propto E_\g^{1/2}$ $(E_\g^{-1})$ before (after) the peak corresponding to the matter-dark energy equality with $E_\g \lesssim m_\f/2$ due to the redshift. This component has an energy width of $\O(10)\%.$ 
We do not consider it since the event rate for this component within a bin is suppressed by the large $R$.

The second term, which is our focus, represents the component from a nearby galaxy $i$, 
\beq
\frac{d\Phi_{\g,i}}{dE_\g }=\int ds d \Omega \frac{e^{-\tau[s, \Omega ] s}}{4\pi s^2}   \left(\frac{\Gamma_{\f}  \rho^{i}_\phi(s,  \Omega )}{m_\phi}\right) \, s^2\, \frac{d N_{\f,i}}{dE}[ s, \Omega] 
\eeq
where $s$ is the line distance of sight, 
$\rho^{i}_\f$ and $\frac{d N_{\f,i} }{dE} [s, \Omega]$ represent the DM density distribution and photon spectrum from a single DM decay around the galaxy $i$, respectively, which both rely on DM halo model and property of the galaxy $i$.  
$\tau $ is the (averaged) optical depth. We take 
\beq \tau\simeq 0\eeq
{unless otherwise stated, since we focus on infrared photons, which typically have high transparency. One exception is the galactic center of the Milky-way, which we will discuss in Sec.\ref{sec:MW}.}

In particular, the photon spectrum 
 Doppler-shifts as 
\beq
\frac{d N_{\f, i} }{d E} = 2 \int{d^3\vec{v} f_i({v}, s ,\Omega ) \delta(E-(1-\vec{v} \cdot \vec{\Omega})m_\f/2 )}.
\eeq
Here $f_i$ is the DM velocity distribution. 
Throughout this paper,  we assume that the DM velocity has a symmetric gaussian distribution
\beq\laq{fdis}
f_i= \prod_{a=1,2,3} \frac{1 }{\sqrt{2\pi}\s_i  } e^{- \frac{[\vec{v}-\vec{v}_{i}]_a^2 }{2\s_i^2} }
\eeq
and $\s_i$ is the one-dimensional velocity dispersion and $\vec v_i$ is the average velocity of the galaxy $i$ in the laboratory frame. 
Then we obtain 
\beq
\frac{d N_{\f,i}}{dE}\simeq \frac{4}{m_\f}  \frac{1 }{\sqrt{2\pi}\s_i  } e^{- \frac{\(\frac{2E}{m_\f}-1+\vec{v}_i\cdot \vec{\Omega} \)^2 }{2\s_i ^2} }.
\eeq
When $\s_i$ is small and when the size of the galaxy spreads in a small angle we can use the form
\beq
\lim_{\s_i\to 0}\frac{d N_{\f,i}}{dE} \to \frac{4}{m_\f} \delta{(\frac{2E}{m_\f}-1 +\vec{v}_i \cdot \vec\Omega )} \approx \frac{4}{m_\f} \delta{(\frac{2E}{m_\f}-1 +v_i^r)}
\eeq
where we have defined the radial velocity of the galaxy $i$ by 
 $v^r_i\equiv \vec{\Omega}_i\cdot \vec{v}_i (\ll 1)$ with $\vec{\Omega}_i$ being the 
the direction of the center of galaxy $i$. 
In the following, we neglect the contribution of $\vec{v}_i$ unless otherwise stated. 

In general, the spectrum (from nearby DM decay) has a peak around $E_\g \approx m_\f/2$.
The flux satisfies $\left.\frac{d \F_{\g, i}}{d E_\g }\right|_{E_\g/m_\f={\rm const}} \propto g_{\f\g\g}^2 m_\f,$ when $E_\g/m_\f$ and the DM distribution are fixed.

\subsection{Photon spectrum from dwarf spheroidal galaxies.}
The dSphs around the Milky Way are known to be the most extreme DM-dominated objects, e.g., \cite{Mateo:1998wg, Kleyna:2001us}. 
They are good candidates to search for the eV DM not only because of the larger central mass to light ratios $\sim \O(10-100)$ but also because of the small velocity dispersion, $\s \lesssim 10$ km/s. 
Even for the spectral resolution of $R=10^{3-5}$, the photons from the DM decay are detected in a single bin or spread over only a few bins.\footnote{In this sense, a spectrograph with $R\sim 10^{4}-10^{5}$ like the WINERED happens to be optimized for the DM search. } 
In particular, if we concentrate on the center of the dSph, the velocity dispersion gets even more suppressed (c.f. Fig.\ref{fig:osc}). 
Therefore, we assume that the signal spectrum is line-like in the following spectrographs for simplicity.\footnote{{Our conclusions will not change even if the spectrum spreads over a few bins.}}

The spectrum is given by
\beq
\laq{dsph}
\frac{d^2 \F_{\g,i=\rm dSph}}{d E_\g d \Omega }\simeq \frac{D[\D \Omega]}{4\pi \D \Omega}\frac{2e^{-\t \times { d}} \G_{\f}}{m_\f} \delta{(E- m_\f/2)}
\eeq
here $\D \Omega=2\pi\times (1-\cos[\theta])$  with $\theta$ being the angle measured from the center of galaxy $i$. 
We also defined the so-called D-factor, 
\beq
D[\D \Omega]\equiv \int_{\D \Omega} d \Omega d s \rho^i_{\f} (s, \Omega) . 
\eeq
The D-factor can be obtained from references e.g.~\cite{Combet:2012tt,Geringer-Sameth:2014yza,Bonnivard:2015xpq,Bonnivard:2015tta,Hayashi:2016kcy,Sanders:2016eie,Evans:2016xwx,Hayashi:2018uop,Petac:2018gue}. 
 We translate the data from Refs.\, \cite{Evans:2016xwx} to show $\frac{d^2 \F_{\g,i=\rm dSph}}{ d \Omega }$ in the Table.~\ref{table:Ds} by taking  $\theta=\theta_{\rm max}$ in the reference. We also take the relevant information of the dSphs from  the SIMBAD database~\cite{Wenger:2000sw}.\footnote{https://simbad.unistra.fr/simbad/} 
We expect that the approximation of $\theta \approx \theta_{\rm max}$ gives a conservative estimation, given that the DM halo model may have theoretical uncertainty.
If the dSph has a cusped DM density (e.g. \cite{Read:2018pft, Hayashi:2020jze}), the sensitivity reach is enhanced since $D[\D \Omega ]/\D\Omega$ gets larger with smaller $\D \Omega$  (again c.f. Fig.~\ref{fig:osc}). 
For instance, if we consider the NFW profile~\cite{Navarro:1995iw,Cirelli:2010xx}, $\rho_\f^i\propto 1/|\vec{r}-\vec{r}_i|$ close to the center of galaxy ($\vec r=\vec r_i$). 
Thus at the small angle $\D \Omega \sim \theta^2$, $D[\D \Omega]\propto \theta^2 (1+2{\rm arctanh}[\frac{\theta_0}{\sqrt{\theta^2+\theta_0^2}}])$ with $\theta$ being the angle measured from the center of the galaxy, $i$, and 
 $\theta_0$ corresponds to the cutoff for the cusp behavior. The arctanh correction to the $\theta^2$ scaling
provides a mild enhancement of $\O(10)$ for $D[\Delta \Omega]/\Delta \Omega$ if
 $\theta \sim (10^{-3}-10^{-4}) \theta_0$.\footnote{{On the other hand, in this case, we may need to consider the interstellar absorption well inside the dSph.}} {Note that 
 the WINERED's (NIRSpec, IRCS) angular resolution can resolve $\d \theta\sim 0.0001^\circ$ ($\d \theta\sim 0.00001^\circ$)} while the typical size is $\theta_{\rm max}\sim \O(0.1^{\circ}-1^\circ).$
 Thus the estimation of the photon flux here may be potentially an order of magnitude smaller when we see the center of the galaxy, $i$.

\begin{table*}
\caption{Photon flux from dwarf spheroidal galaxies. The result is calculated from the D-factor listed in \cite{Evans:2016xwx}. 
Relevant information of the dSphs is taken from the SIMBAD database~\cite{Wenger:2000sw}. 
$\theta_{\rm max}$ is the angle between the center of the dSph and the distance to the outermost member star. 
Here Flux$^{\rm norm}$ denotes $\frac{d\Phi_{\g,i=\rm dSph}}{d\Omega} \left(\frac{m_\phi}{2\,{\rm eV}} \frac{g_{\phi\gamma\gamma}} {10^{-10}\,\rm GeV^{-1}}\right)^{-2}$. 
$d \F_{\g,i=\rm dSph} /d \Omega$ is the coefficient of the delta-function in \Eq{dsph} with $\Delta \Omega= 2\pi(1-\cos[\theta_{\rm max}])$. We take the mean free path of photon $1/\tau =\infty.$ 
The dSphs with the superscript $*$ ($\star$) cannot (is difficult to) be seen from the Magellan telescope, which is one of the main targets. 
In this case, the DM search can be possible with a spectrograph in the northern hemisphere or the sky. The DM search can be also performed with the WINERED in 10\,m class telescopes in the northern hemisphere, such as the Subaru telescope, the Keck telescope and the Gemini North telescope. }
\begin{center}
\begin{tabular}{llccccc}
\hline
\hline
Name & $d$ & $\theta_\mathrm{max}$ & $10^6 {\rm Flux}^{\rm norm}$ & Right ascension & Declination& $v_i^r$ \\
& $[\mathrm{kpc}$]& [$^\circ$] &[${\rm cm}^{-2} {\rm s}^{-1} {\rm arcsec}^{-2}$]& & & [km/s]\\
\hline
Carina&$105$ & $1.26$& $1.79^{+  0.86}_{- 0.58 }$ & 06:41:36.7 & -50:57:58& 222.9\\
Draco$^{*}$&$76$ & $1.3$&$ 4.75^{+  1.51 }_{-1.15 }$ & 17:20:14.335 & +57:55:16.39&-291.0 \\
Fornax&$147$ & $2.61$&$ 1.95^{+  0.45 }_{- 0.37 }$ & 02:39:59.3 & -34:26:57& 55.3\\
Leo I&$254$ & $0.45$&$ 4.15^{+  1.58 }_{- 1.14 }$ & 10:08:28.12 & +12:18:23.4& 282.5 \\
Leo II&$233$ & $0.23$&$ 4.69^{+1.49 }_{- 1.13 }$ & 11:13:28.13 & +22:09:10.1& 78.0\\
{Sculptor}&$86$ & $1.94$&$ 2.75^{+1.13 }_{- 0.80 }$ &01:00:09.401 &{-33:42:32.00}&  111.4\\
Sextans&$86$ & $1.7$&$ 2.42^{+1.00 }_{- 0.71 }$ & 10:13:02.9 & -01:36:53& 224.2\\
Ursa Minor$^{*}$&$76$ & $1.37$&$ 4.48^{+1.42 }_{- 1.08}$ &15:09:11.34 & +67:12:51.7& -246.9\\
%\hline
Bo\"otes I&$66$ & $0.47$&$ 0.98^{+1.06 }_{- 0.35 }$ & 14:00:00.0 & +14:30:00&99.0\\
Coma Berenices&$44$ & $0.31$&$ 8.54^{+3.81 }_{- 2.63}$ & 12:26:59.0 & +23:54:15&  98.1\\
Canes Venatici I$^{\star}$&$218$ & $0.53$&$ 2.79^{+0.34 }_{- 0.30 }$ & 13:28:03.5 & +33:33:21&  30.9\\
Canes Venatici II$^{\star}$&$160$ & $0.13$&$ 5.84^{+3.42 }_{- 2.16 }$ &12:57:09.6 & +34:19:12& -128.9 \\
Hercules&$133$ & $0.28$&$ 1.95^{+1.29 }_{- 0.77 }$ & 16:31:02.0 & +12:47:30& 45.2\\
Leo IV&$154$ & $0.16$&$ 2.27^{+4.13 }_{- 1.46 }$ & 11:32:57.0 & -00:32:00& 132.3\\
Leo V&$178$ & $0.07$&$ 5.55^{+13.25 }_{- 3.07}$ & 11:31:09.60 & +02:13:12.0 & 173.3\\
Leo T&$417$ & $0.08$&$ 9.51^{+5.22 }_{- 3.37 }$ & 09:34:53.4 & +17:03:05&  38.1\\
Segue 1&$23$ & $0.35$&$ 10.62^{+6.21 }_{- 3.92 }$ & 10:07:03.2 & +16:04:25& 208.5\\
Segue 2&$35$ & $0.19$&$ 1.47^{+2.48 }_{- 1.27 }$ & 02:19:16.0 & +20:10:31&-39.2\\
Ursa Major I$^{*}$&$97$ & $0.43$&$ 7.9^{+2.76 }_{- 2.04 }$ & 10:34:52.80 & +51:55:12.0& -55.3 \\
Ursa Major II$^{*}$&$32$ & $0.53$&$ 13.99^{+7.68 }_{- 4.96}$ & 08:51:30.0 & +63:07:48& -116.5\\
Willman 1$^{*}$&$38$ & $0.5$&$ 5.2^{+9.46 }_{- 2.71 }$ & 10:49:21.0 & +51:03:00& -12.3\\
Reticulum II&$30$ & $1.0$&$ 1.88^{+3.07 }_{- 0.61}$ & 03:35:42.1 & -54:02:57& 62.8\\
Tucana II&$57$ & $1.0$&$ 7.49^{+13.13 }_{- 3.65 }$ & 22:51:55.1 & -58:34:08 & -129.1\\
Horologium I&$79$ & $0.5$&$ 2.93^{+5.71 }_{- 1.08 }$ & 02:55:31.7 & -54:07:08& 112.8 \\
Hydra II&$134$ & $0.5$&$ 0.38^{+0.66 }_{- 0.33 }$ & 12:21:42.1 & -31:59:07& 303.1\\
Pisces II&$182$ & $0.5$&$ 1.25^{+3.39 }_{- 0.75 }$ & 22:58:31.0 & +05:57:09& -188\\
Grus I&$120$ & $0.5$&$ 1.89^{+3.56 }_{- 1.68 }$ & 22:56:42.4 & -50:09:48 &  -140.5\\
\hline
\end{tabular}

\end{center}
\label{table:Ds}
\end{table*}
\subsection{Signal-to-noise ratio (SNR)}

During the history of the Universe, cumulative photons from the early galaxies compose the EBL
\cite{Matsumoto:2010pz, Windhorst:2010ib, Kashlinsky:2012zz,  Mitchell-Wynne:2015rha, Seo:2015fga, Ahnen:2016gog, Helgason:2016xoc, HESS:2017vis, Fermi-LAT:2018lqt, Desai_2019, Lauer:2022fgc} (see also Ref.~\cite{CTAConsortium:2017dvg} for the future reaches in Cherenkov Telescope Array.).
We note that the flux is $10^{-4}-10^{-3}$ nW cm$^{-2}$sr$^{-1}$ at $\lambda=\O(1) \m$m measured from the TeV gamma-ray annihilation with the EBL.  A more important background is the zodiacal light, caused by the sunlight scattered by the interplanetary dust.
 This usually dominates the sky background in the energy range that we are focusing on. 
For instance in the south ecliptic pole it is $\text{BG}_{\rm flux}\sim 10^{-2}-1$ nW cm$^{-2}$sr$^{-1}$ (see e.g. Ref.\,\cite{leinert19981997}). There are also other sources, like the star light, which will be also included.
As we have mentioned, a high resolution $R\gg 1 $ and angular resolution of $\O(\rm arcsec^{2})$ suppresses those backgrounds made up of the continuous spectrum. 
For instance, we can see that the background photons in a single bin scales as 
\beq
\D \lambda \frac{\partial^2 \F_\g^{\rm zodiacal}}{\partial \lambda \partial \Omega}\sim 10^{-7}-10^{-6}  {\rm cm}^{-2} {\rm s}^{-1} {\rm arcsec}^{-2} \frac{28000}{R}.
\eeq
Thus a large $R$ suppresses the sky background noise. 
Interestingly, this can be smaller than the typical signal flux from the dSphs (see Table.\,\ref{table:Ds}) with the ALP DM around the astrophysical bound for a large enough $R\sim 10^{3-4}$.

There are also backgrounds depending on the environment of the experiments and detectors. 
Here we perform this environment-dependent background estimation by using two examples, WINERED at the Magellan telescope on Earth and NIRSpec at the JWST in the $L_2$ Lagrange point.

\paragraph{CASE of WINERED on Earth}
Here let us take the Magellan Clay $6.5$m telescope as an example to search for the ALP DM. We can use the WINERED without any modification to search for the ALP DM by looking at certain dSphs. 

First of all, the Magellan  telescope can only see the galaxy in Declination $40^\circ$ to $-90^\circ$. 
Therefore it is hard to see Draco, Ursa Minor, Ursa Major I, Ursa Major II, and Willman 1 in Table. \ref{table:Ds}. Secondly, the sensitive regime in ground-based telescopes suffers from atmospheric absorption. The resulting sensitive wavelengths are in
\beq \laq{range} 0.91-0.94,0.96-1.10, \AND 1.14-1.35\m \rm m.\eeq Otherwise, there would be absorption by water in the atmosphere. 

Keeping those in mind, the SNR is evaluated as follows. 
The SNR is defined in the form of
\beq
\laq{SNR}
\text{SNR} = \frac{I_\text{signal}}{\sigma_\text{noise}}.
\eeq
Here $I_{\text{signal}}$ is the number of the signal events 
\begin{align}
I_\text{signal} &=  \int^{\Delta\omega}{d\Omega \frac{d\Phi_{\g,i=\rm dSph}}{d\Omega}\ S_\text{tel}  \ t_\text{exp} \ \eta_\text{ins} \ \eta_\text{atm}}\\
&\sim  \frac{d\Phi_{\g,i=\rm dSph}}{d\Omega}  {\pi \over 4}D_\text{tel}^2 \ W_\text{slit} \ L_\text{slit} \ t_\text{exp} \ \eta_\text{ins} \ \eta_\text{atm}
\end{align}
which is not suppressed by $R$ because we focus on the line-like spectrum.
The flux of signal photons from each dSph can be found in the fourth column of Table. \ref{table:Ds}. The other parameters, including the spectrograph-specific parameters, are explained in Table. \ref{table:WINERED}. 
Here we assume that the integrant does not change in the slit angle, $\Delta \omega $, in the last approximation. 
 $t_\text{exp}$ is the total exposure duration of the observation, e.g. single night observation (around 4h) implies 
$
t_{\text{exp}}= 14400{\rm sec}.
$
For instance 
\beq
I_{\text{signal}}\approx 4000 \frac{t_{\text{ext}}}{14400\rm sec} \frac{d\F_{\g,i=\rm dSph}/d\Omega}{10^{-6} \text{cm}^{-2}\text{s}^{-1}\text{arcsec}^{-2}}.
\eeq

The total background noise, 
\beq
\sigma_\text{noise} = \sqrt{\sigma_\text{photon}^2 + \sigma_\text{readout}^2 + \sigma_\text{thback}^2 + \sigma_\text{sky}^2},
\eeq
is estimated as follows: 
\begin{align*}
\sigma_\text{photon} &= \sqrt{I_\text{signal}}\\
\sigma_\text{readout} &= \sigma_{\text{readout}_\text{pix}} \sqrt{N_\text{readout} {W_\text{slit} \ L_\text{slit} \over l_\text{pix}^2}}\\
\sigma_\text{thback} &= \sqrt{P_\text{thback} \ {W_\text{slit} \ L_\text{slit} \over l_\text{pix}^2} \ t_\text{exp} \ N_\text{sky}}\\
\sigma_\text{sky} &= \sqrt{(P^{\rm atm}_\text{sky}+P^{\rm natm}_\text{sky}) \ W_\text{slit} \ L_\text{slit} \ \ t_\text{exp} \ N_\text{sky}\ \eta_{\rm ins}}
\end{align*}
Again their definitions, as well as the experimentally measured values, are summarized in Table.\,\ref{table:WINERED}. In particular we note that
$P^{\rm natm}_{\rm sky}$ is important in $\s_{\rm sky}.$ It depends on direction of the target dSph. 
We will evaluate this background following the IPAC IRSA model (see e.g. Refs.\,\cite{Arendt:1998aj, Schlegel:1997yv, 1997IPAC, Zubko:2003eg, 1998IPAC,Brandt:2011ka}).\footnote{https://irsa.ipac.caltech.edu/applications/BackgroundModel/}
For instance, we can use the sky background flux from Segue 1 as $\frac{\partial^2 \Omega_{\gamma}^{\rm sky}}{\partial \Omega \partial \lambda}= {\rm 0.55593 Mjy/sr} $ (including 82\% zodiacal light, 17\% of starlight, and others) at  $\lambda=1\m$m on 30th Apr, 2023.
We can easily find that for a single night (4h) observation, we obtain 
\begin{align}
\s_{\rm readout}\approx 190,\s_{\rm thback}\approx 480, \s^{\rm Segue1}_{\rm sky}\approx 79.
\end{align}
Thus the dominant noise is from the thermal radiation of photons in the environment. This is still well below the signal events for not too small $g_{\f\g\g}$ thanks to the ``warm" instrument of the WINERED \cite{2022WINERED}. We also note that in this case ``WIDE" mode with $R=28000$ is already good enough for suppressing the background. If the thermal background can be further reduced, ``HIRES-Y" and ``HIRES-J" modes can be more efficient in detecting DM. 
The declinations of Canes Venatici I  and Canes Venatici II in Table. \ref{table:Ds}  indicate to have a thick atmosphere in the line of sight. Our $\sigma_{\rm sky}$ should be underestimated in those cases. 
On the other hand,
the sky background depends on time especially due to the zodiacal light. 
However, $\s_{\rm sky}$ is subdominant unless $\frac{\partial^2 \Omega_{\gamma}^{\rm sky}}{\partial \Omega \partial \lambda}$ is 1-2 orders of magnitude larger than what we have estimated. 
 Let us suppose that the measurement is performed at the season when the zodiacal light is small and we do not consider  Canes Venatici I  and Canes Venatici II for conservativeness. Then we are allowed to neglect $\s_{\rm sky}$, which is smaller than the other noises.

\begin{table}
\caption{The parameters list for the SNR evaluation~\cite{ikeda2006winered, yasui2008warm, kondo2015warm, ikeda2016high, ikeda2018very, 2022WINERED}. In this table, we assume a photon wavelength of $1.22\m$m, but we consider they do not change much in the range of \Eq{range}. We also consider the ``WIDE" mode $(R=28000).$ }
\begin{tabular}{ccl}
\hline
\hline
parameter		&	value		&	description\\
\hline
$\sigma_\text{photon}$	& see the text	& 	 noise for photons from the signal\\
$\sigma_\text{readnoise}$	& see the text	&	{read noise  from  the focal plane array}\\
$\sigma_\text{thback}$	&  see the text	&	{noise for  background photons}\\
$\sigma_\text{sky}$	&	 see the text&	{noise from the sky background light}\\
\hline
$S_\text{tel}$	&	${\pi \over 4} D_\text{tel}^2[\text{cm}^2]$	&	the entrance aperture area of the Magellan telescope	\\

$D_\text{tel}$	&	650[\text{cm}]	&	the diameter of the Magellan telescope\\
$\Delta \omega$	&	$W_\text{slit} L_\text{slit}[\text{arcsec}^2]$	&	the solid angle the WINERED observes through the slit\\
$W_\text{slit}$	&	0.3[arcsec]	&	the slit width of the WINERED\\
$L_\text{slit}$	&	6[arcsec]	&	the slit length of the WINERED\\
$l_\text{pix}$	&	0.15[arcsec/pix]	&	the pixel scale, the angle a pixel spans on the sky	\\
$t_\text{single}$	&	900[sec]	&	the longest exposure duration of a single exposure	\\
$N_\text{readout}$	&	$t_\text{exp} \over t_\text{single}$	&	the number of readouts	\\
$\sigma_{\text{readout}_\text{pix}}$	&	5.3[$e^-$]	&	the readout noise of each pixel	\\
$\eta_\text{ins}$	&	0.51	&	{the instrumental throughput of ``WIDE" mode}\\
$\eta_{\text{atm}}$	&	0.9-0.95&	the atmospheric absorption ratio for around $\lambda_c$\\
$P_\text{thback}$	&	 0.1[$\gamma$/sec/pix]	&	the background photon radiation for $T=290[\text{K}]$\\
$P^{\rm atm}_\text{sky}$	&	0.128[$\gamma$/sec/arcsec$^2$]	&	the radiation from the atmosphere of 15.9[mag] at J-band\\
$P^{\rm natm}_\text{sky}$	&	see the text	&the sky background radiation not from the atmosphere.\\
$N_\text{sky}$	&	2	&	the number of the sky frames	\\
\end{tabular}
\label{table:WINERED}
\end{table}

\begin{figure}[!t]
\begin{center}  
   \includegraphics[width=160mm]{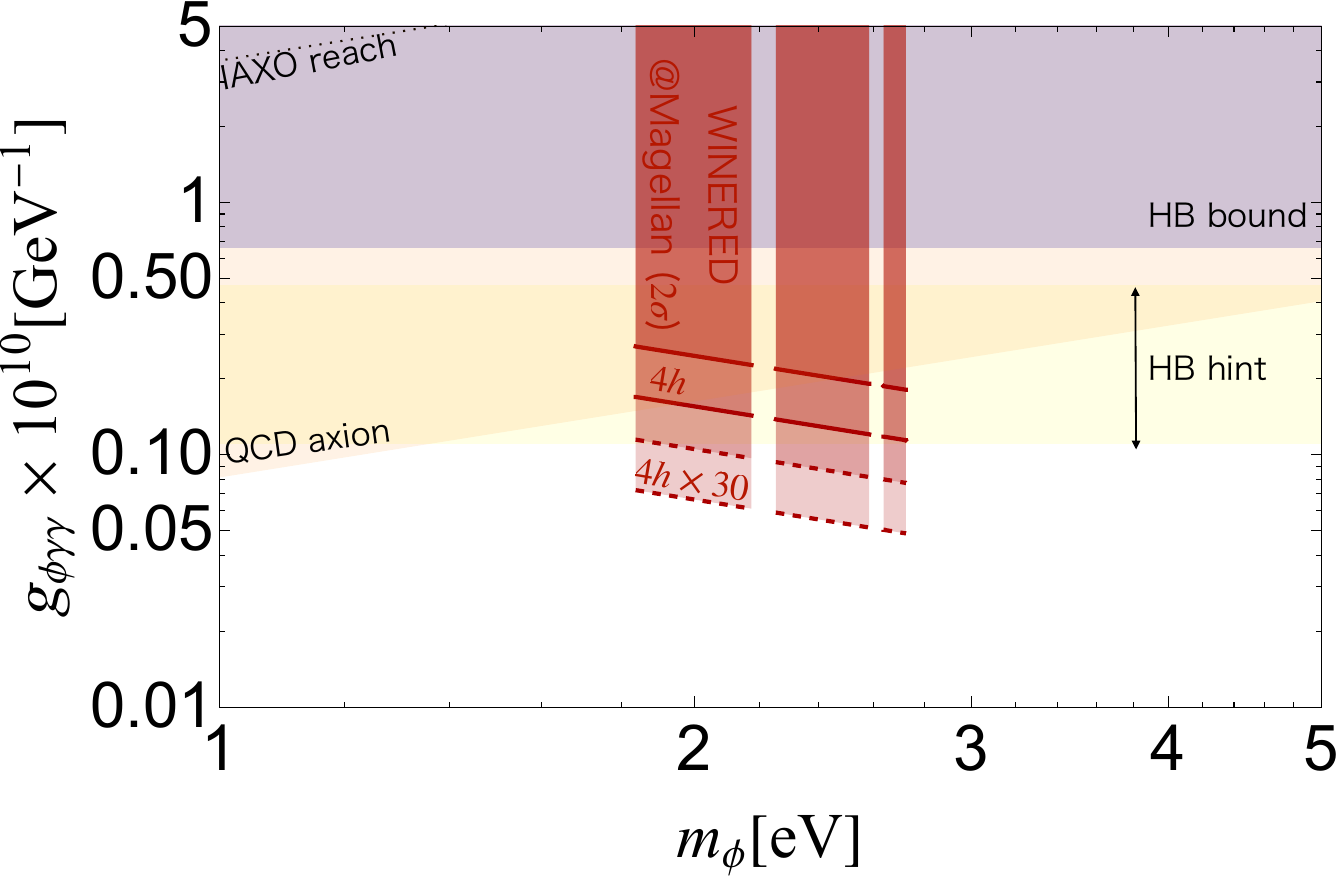}
      \end{center}
\caption{
The sensitivity reaches of WINERED at Magellan Clay 6.5m telescope by observing the ALP DM in Segue 1 in $m_\f-g_{\f \g\g}$ plane. The $2\s$ reach with a night and 30 nights,  i.e. 4h and 4h $\times 30$, observations
 are
 shown in red solid and dashed lines,  respectively. 
The two solid/dashed lines take account of the uncertainty of the DM halo (see Table.~\ref{table:Ds}.). 
Also shown are the cooling bound of the horizontal branch star~\cite{Raffelt:1985nk,Raffelt:1987yu,Raffelt:1996wa, Ayala:2014pea, Straniero:2015nvc, Giannotti:2015kwo, Carenza:2020zil} and the hint as well as the IAXO reach and QCD axion prediction~\eq{QCDaxion}. 
} \label{fig:para}
\end{figure}

 We display the future reach of the ALP DM in $m_\phi-g_{\phi\gamma \gamma}$ plane in Fig.\ref{fig:para} by looking at the Segue 1
from Magellan Clay 6.5m telescope by using the WINERED detector of ``WIDE" mode, i.e. $R=28000$, for 4 (120) hours in the red solid (dashed) lines. 
Also shown are the cooling bound (hint) from the horizontal branch (HB) star~\cite{Raffelt:1985nk,Raffelt:1987yu,Raffelt:1996wa, Ayala:2014pea, Straniero:2015nvc, Giannotti:2015kwo, Carenza:2020zil}, the reach of IAXO~\cite{Irastorza:2011gs, Armengaud:2014gea, Armengaud:2019uso, Abeln:2020ywv} and the QCD axion prediction \eq{QCDaxion}.

\paragraph{Case of NIRSpec in the Sky}
The spectral resolution of the NIRSpec at the JWST is $R=100,1000,2700$ with different Disperser-filter combinations.

By using the parameters in Table.~\ref{table:JWST},
the signal events can be evaluated as
\beq
I_{\text{signal}}\approx 1300 \frac{t_{\text{ext}}}{14400[\rm sec]} \frac{d\F_{\g,i=\rm dSph}/d\Omega}{10^{-6} \text{cm}^{-2}\text{s}^{-1}\text{arcsec}^{-2}}.
\eeq
Again it does not depend on $R$. 
Since we do not have the atmosphere thermal radiation, 
 the sky background is estimated 
\beq 
\s_{\rm sky}\simeq \sqrt{ \frac{\partial^2 \F_{\gamma}^{\rm sky}}{\partial \Omega \partial \lambda}   \D \l  {\pi \over 4}  D_\text{tel}^2 \ W_\text{slit} \ L_\text{slit} \ t_\text{exp} \ \eta_{\rm ins} C_{\rm det} }.
\eeq
Here $C_{\rm det}$ depends on the detection strategy where we take $C_{\rm det}=2$ in the following. 
For instance, from Segue 1 we can find the sky background flux, $\frac{\partial^2 \F_{\gamma}^{\rm sky}}{\partial \Omega \partial \lambda}= {\rm 0.40 Mjy/sr} $ (including 63\% zodiacal light, 35\% of starlight, and others) at  $\lambda=2\m$m on 30th Apr, 2025 at $L_2$ Lagrange point. Then we obtain 
\beq
\sigma^{\rm Segue 1}_{\rm sky} \sim 140 \sqrt{\frac{t_{\rm ext}}{14400[\rm sec]} \frac{2700}{R}}. 
\eeq
We note that in the target wavelength (the high-resolution mode) $0.7-5.3\m$m, this does not change by more than an order of magnitude. Thus we do not introduce the wavelength dependence in this paper. 

On the contrary, we neglect $\sigma_{\text{thback}}$ 
\beq
\sigma_{\text{thback}} \sim 0 
\eeq
since the temperature at the cold side of the JWST is $\lesssim 50$K and the thermal radiation should be suppressed.\footnote{See also https://jwst-docs.stsci.edu/jwst-general-support/jwst-background-model} 
{This, as well as the absence of atmospheric thermal radiation, is the difference from infrared spectrographs on Earth.}

We also estimate the readout noise from the parameters we assumed in Table. \ref{table:JWST}.
\begin{align}
\s_{\rm readout}\approx 97\sqrt{\frac{t_{\text{ext}}}{14400[\rm sec]}}. 
\end{align}
{This may be slightly different from the realistic case in the JWST, and thus, we consider our analysis as a toy model for the background estimation.}
 That said, we have checked that for certain $m_\f, g_{\f \g\g}$ the SNR derived by us is not very different from that derived from the JWST exposure time calculator.\footnote{https://jwst.etc.stsci.edu} 
Now we are ready to evaluate \Eq{SNR}. 

\begin{figure}[!t]
\begin{center}  
   \includegraphics[width=160mm]{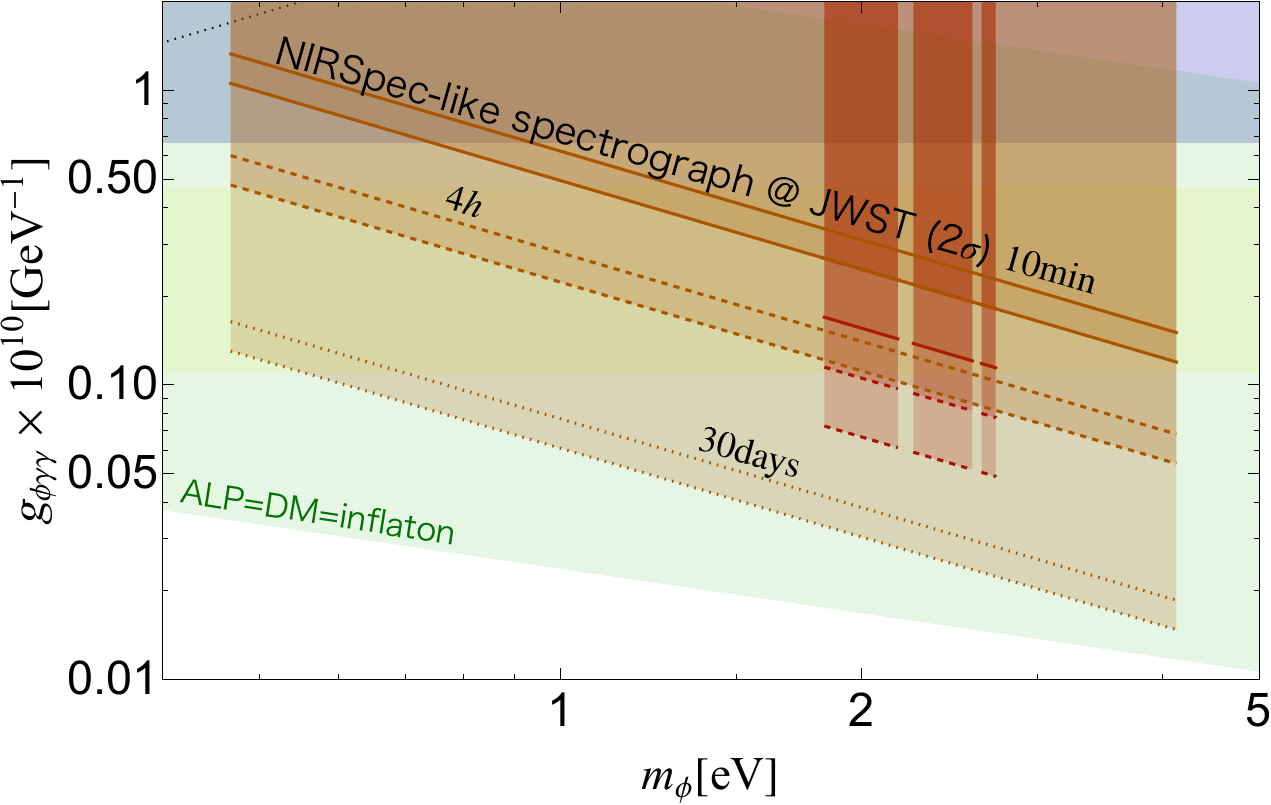}
      \end{center}
\caption{
Sensitivity reaches of a NIRSpec-like detector measuring Segue 1 at the JWST. 
The  $2\s$ reaches with 10min, 4h,  and 30 days ($30 \times 24$h) observations are shown in red solid, dashed, and dotted lines, respectively. Others are same in Fig. \ref{fig:para} except that we also show the parameter region of inflaton=DM=ALP scenarios \eq{ALP} with $c_\g=0.1-10$. We did not show the QCD axion regime just for the simplification of the figure, but it is also searchable. 
The WINERED reach in the previous figure is also shown for comparison. 
} \label{fig:para2}
\end{figure}

Again let us consider Segue 1 dSph. We display the future reach of the ALP DM in $m_\phi-g_{\phi\gamma \gamma}$ plane in Fig.\ref{fig:para2} by using the NIRSpec-like spectrograph defined by the parameter in the Table \ref{table:JWST} as the detector at JWST . The reach for 1/6 (4, $24\times 30$) hours of exposure is shown by the orange solid (dashed, dotted) lines. The two lines denote the uncertainty in the D-factor. 
Here we assume the wavelength range of $0.7-5.3\m$m, which is the same as the combined range of F100LP, F170LP, and F290LP,
 and  $R=2700$. Also shown is the prediction of the hypothesis, ALP$=$DM$=$inflaton \eq{ALP}. The other regions are the same as in the previous Fig.\ref{fig:para}, but the QCD axion region is not shown just for simplification of the figure.

\begin{table}
\caption{The relevant parameters for a NIRSpec-like spectrograph at JWST. 
See also Table\ref{table:WINERED} for description.}
\begin{center}
\begin{tabular}{cc}
parameter		&	value		\\ 
\hline
$S_\text{tel}$	&	${\pi \over 4} D_\text{tel}^2[\text{cm}^2]$	\\
$D_\text{tel}$	&	650[\text{cm}]	\\
$\Delta \omega$	&	$W_\text{slit} L_\text{slit}[\text{arcsec}^2]$	\\
$W_\text{slit}$	&	0.2[arcsec]	\\
$L_\text{slit}$	&	3.3[arcsec]	\\
$l_\text{pix}$	&	0.1[arcsec/pix]	\\
$t_\text{single}$	&	3600[sec]	\\
$N_\text{readout}$	&	$t_\text{exp} \over t_\text{single}$	\\
$\sigma_{\text{readout}_\text{pix}}$	&	6[$e^-$]	\\
$\eta_\text{ins}$	&	0.4	\\
\end{tabular}
\end{center}
\label{table:JWST}
\end{table}

\section{Other approaches with state-of-art spectrographs}
\label{sec:comp}
Here we denote some complementary approaches by taking advantage of the high angular and spectral resolutions of the state-of-art spectrographs.
We consider they should be useful in the future for discovering DM.
\subsection{Indirect detection from the Milky Way galaxy center}
\label{sec:MW}
Let us also discuss the indirect detection of the infrared DM from our Milky Way galaxy. {As we will see, it is more challenging due to the larger velocity dispersion of the DM and the bright background as well as interstellar absorption.}

\paragraph{Velocity dispersion around the Milky Way galactic center}
The typical velocity dispersion of the DM, $\O(10^{-3})$, in our galaxy implies that the photon ``line" is broadened with a width of $\O(10^{-3}).$
To discuss the effect of the velocity dispersion in more detail, 
we further assume that $\s[r]$ in \Eq{fdis} has a spherical distribution around the Milky Way galactic center. Here $r$ is the radial distance from the galactic center. 
According to the Jeans equation, the one-dimensional velocity dispersion is calculated as~\cite{Robertson:2009bh}
\beq
\s^2[r]=\frac{1}{\rho_\f [r]} \int^{r}_{\infty}{ \rho_\f [r'] \frac{d \F_{\rm grav} }{d r' } dr'}
\eeq
with $\F_{\rm grav}$ being the gravitational potential from the galaxy.

\begin{figure}[!t]
\begin{center}  
   \includegraphics[width=160mm]{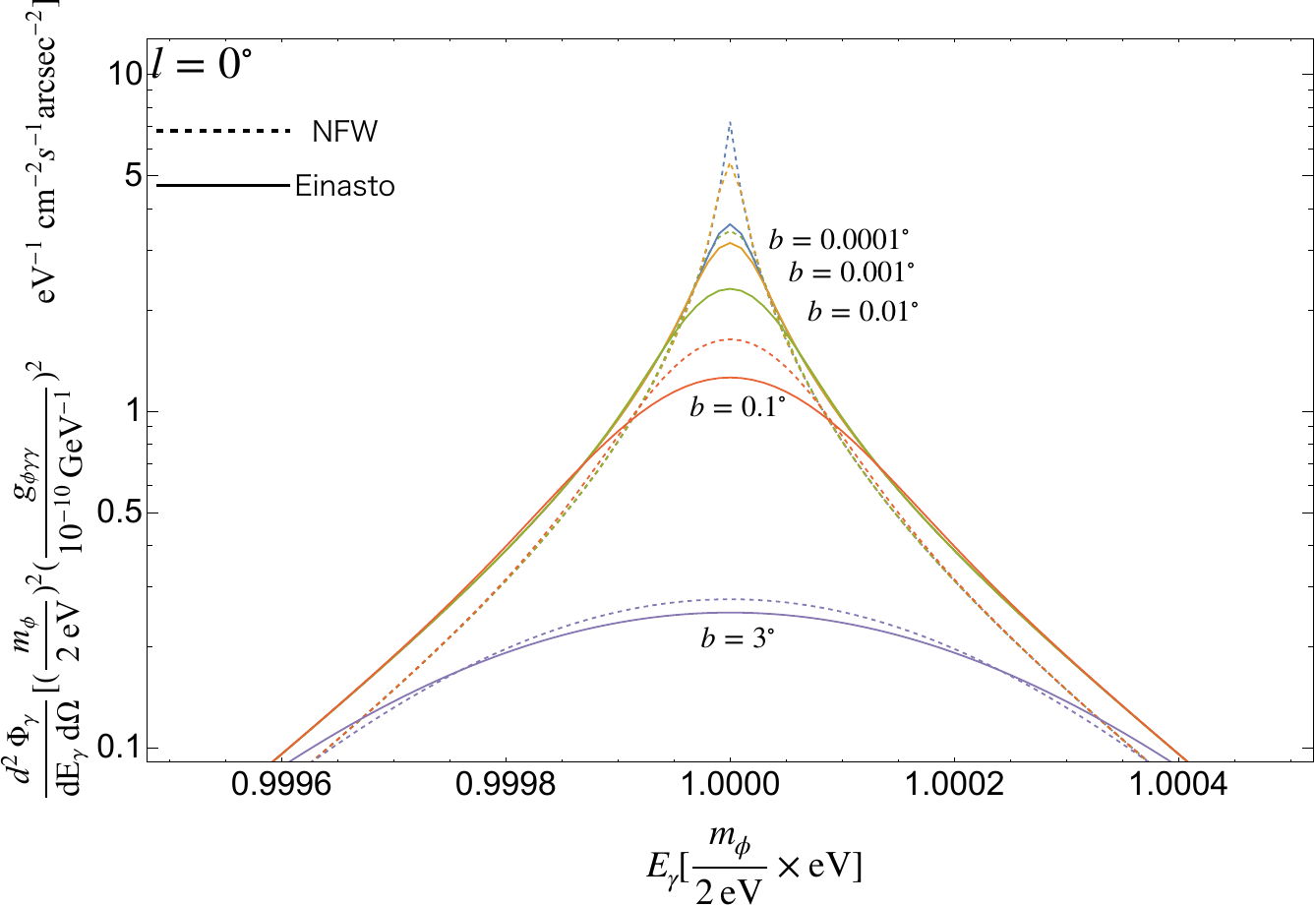}
      \end{center}
\caption{
Angular dependence of the spectrum of infrared light from the ALP DM decay around the center of the Milky Way galaxy. $m_\f=2\EV, g_{\f \g\g}=10^{-10}\GEV$ and we assume the NFW and Einasto profiles in the dashed and solid lines, respectively. The galactic latitude is taken $b=0.0001^\circ, 0.001^\circ,0.01^\circ,0.1^\circ,3^\circ$ from top to bottom. The  longitude is fixed $l=0$. For simplicity, we did not include the Doppler shift effect due to the relative motion of the laboratory, and we set the optical depth $\tau=0$. The interstellar absorption will be taken into account in the text. } \label{fig:osc}
\end{figure}

The angular dependence of the photon spectrum from the Milky Way galaxy is numerically estimated in Fig.\,\ref{fig:osc}. 
We adopt the NFW (dashed line) and Einasto (solid line) profiles given in \cite{Cirelli:2010xx}. For instance, if we adopt the cored Burkert profile~\cite{Burkert:1995yz, Salucci:2000ps} (see also \cite{Salucci:2018hqu}), the $E_\g$ distribution would be flatter. 
The dispersion effect depends on the galactic latitude, $b$, and longitude $l$. 
Interestingly, when the galactic latitude and longitude satisfy $b, l\simeq 0 $, i.e., the line of sight is close to the galactic center, 
the spectrum is close to a line-like one. This is because most of the photons are from the galactic center, where the velocity dispersion is small, and the Doppler effect is suppressed. 
{However, around the center, there is background infrared light from numerous ordinary stars.  }
A spectrograph with a very good angular resolution may find such an around-center direction that the stars are sparse and the light is suppressed.

\paragraph{Intersteller absorption}
Here, we assume that we can look at the direction without a bright object around the galaxy center of the Milky Way. 
Even in this case, in addition to the previous noise, we cannot neglect the optical depth contribution $\tau \neq 0$ due to the scattering of light off of dust grains in the interstellar medium. The interstellar absorption would lead to an extra suppression $ 10^{-0.4 \times A}$ of the flux reaching us.  For instance, the absorption from the galactic center can be characterized by the  $A{(\lambda\sim 2.2\m{\rm m})}\sim 2.7$ [K-band] and $A(\lambda\sim 1.2{\m}\rm m)\sim 8.2$ [J-band]\cite{2009ApJ...696.1407N}. See also Ref.\,~\cite{2020arXiv200401338D} for the $\lambda$ dependence. 
 For the photon corresponding to $m_\f\sim 2\EV$, the suppression is $10^{-0.4 \times A}\sim 10^{-3}$ but for a $m_\f \sim 1\EV $ this is alleviated to $10^{-0.4 \times A}\sim 0.1$. 

In summary, Milky Way indirect detection may be possible by a spectrograph with a relatively low target wavelength and good spectral and angular resolution.  
 One of the candidates is the IRCS\footnote{It should be also useful to search for the ALP DM in the dSphs in the mass range of $0.4-2.7\EV$. The background estimation in this paper can be applied.  } at 8.2m Subaru telescope~\cite{tokunaga1998infrared,kobayashi2000ircs}. The spectral and angular resolutions of the IRCS   have $R=20000$ and $\lesssim 0.01{\rm arcsec}^2$, respectively. The focus of the wavelength is $0.9-5.6 \m$m.  
  From Fig.\ref{fig:osc}, e.g., with $b\sim 0.01^\circ$ $m_\f=1\EV$  and $R\sim 10^{-4}$ we can have a event signal rate in one bin of $10^{-6} \g \rm cm^{-2} s^{-1} arcsec^{-2}\(\frac{ g_{\f\g\g}}{10^{-10}\GEV}\)^2$. 
Detecting this is somewhat more difficult than detecting photons from dSphs but not impossible, and the reach may exceed the cooling bound of horizontal branch stars.

\subsection{Galaxy clusters may not be important}
Other candidates for the conventional indirect detection searches are the galaxy clusters. However, the advantage of the high $R$ detector is lost because of the large velocity dispersion of the DM, $\s \sim 10^{-2}.$
This broadens the signal photon spectrum significantly. Thus we do not consider the galaxy clusters.

 \subsection{Discovering DM via Doppler shift}
{As we have seen that the Doppler shift of the photon energy due to the DM motion plays an important role for the WINERED, NIRSpec or other spectrographs with high spectral resolution in the indirect detection of a DM.}
This is the reason that we focused on the (ultra-faint/classical) dSphs in the measurement due to their small velocity dispersion. 
Conversely, the motion of the dSphs can shift the energy of the photon line while the continuous background lines do not change much. 

\begin{figure}[!t]
\begin{center}  
   \includegraphics[width=160mm]{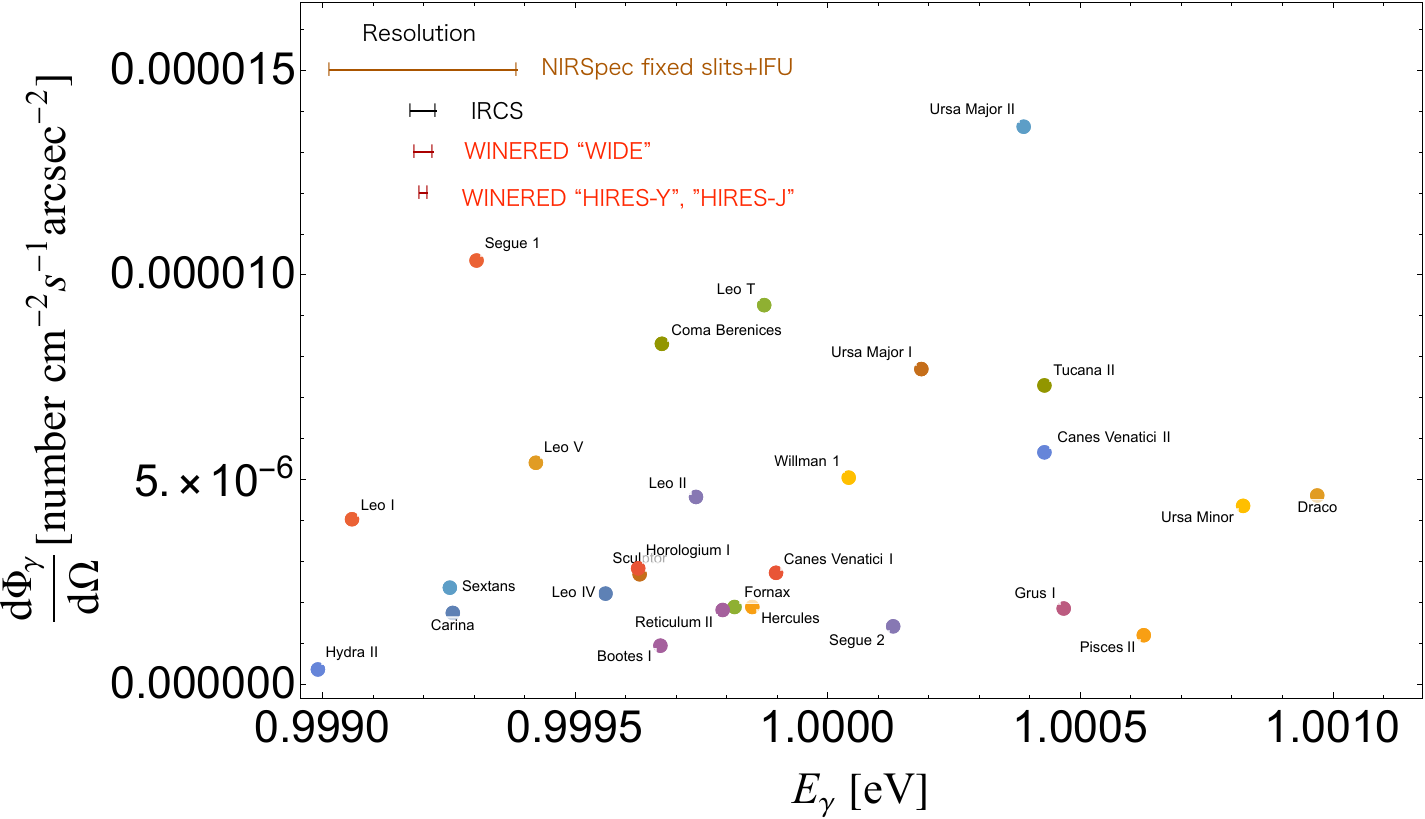}
      \end{center}
\caption{
The photon spectrum in various dSphs in $E_\g$ and $d \F_\g/d\Omega$ plane. 
Here we assume $m_\f=2\EV, g_{\f\g\g}=10^{-10}\GEV^{-1}$, but one can easily obtain the other choices with the certain scaling discussed in the paper. Also shown are the energy resolution of NIRSpec, 
IRCS and WINERED 
in the  ``WIDE" mode and ``HIRES-Y"/``HIRES-J" mode by the left top lines from top to bottom. 
 } 
 \label{fig:redshift}
\end{figure}

We display Fig.\ref{fig:redshift} by including the Doppler shift effect caused by the radial velocities of the dSphs (see Table.\ref{table:Ds}). 
 We fix $m_\f=2\EV, g_{\f\g\g}=10^{-10}\GEV^{-1}$ again. As we have mentioned, the value in the vertical axis scales with $g_{\f\g\g}^2 m_\f^2$ and that in the horizontal axis scales with $m_\f.$ The highest energy resolution of the NIRSpec at JWST, IRCS at Subaru telescope, and WINERED are also shown in the top left lines from top to bottom. 
 One can see that the spectrum of the decay photons originating from different dSphs can be partially (almost) resolved by the NIRSpec (WINERED and IRCS). 
 By fitting the energy difference vs. the known velocity of the dSphs, one can further enhance the sensitivity reaches toward the discovery of the eV DM. Measuring this Doppler shift should be a smoking-gun signal of the monochromatic photons from the DM. 

\section{Conclusions and discussion}

The origin of dark matter (DM) is a long-standing mystery. Paying attention to the recent development of the measurement technology of the infrared photons, we studied the indirect detection of eV DM decaying into a pair of particles, including a monochromatic photon, via infrared spectroscopy. 
Conventionally infrared light suffers from a significant sky background as well as the environment-dependent thermal background.    
However, the background signal, due to the continuous spectrum, gets suppressed with a good spectral resolution in a single bin, while the line-like DM signal does not depend on the spectral resolution unless it is too good. 
We found that this property, as well as the recent development suppressing the thermal background of state-of-art spectrographs, makes it possible to probe the ALP DM in the eV range with the photon coupling beyond the astrophysical bound within a few hours of exposure of a dSph.

Thus we propose to use WINERED, NIRSpec, and IRCS, as well as other high-resolution spectrographs, to perform the indirect detection search of the eV DM by measuring dSphs.  
{In particular, it should be important to take a one-night observation of a dSph by using the WINERED installed at the 6.5m Magellan telescope as the first step of the indirect detection of the eV DM via infrared spectroscopy.
}
\\

Our analysis can be easily extended to search for the dark matter annihilation to the line-like photon.
The good angular resolution may also probe the extended objects decaying into line-like photons. 
In addition, the good spectral resolution may also be useful in precisely measuring dark radiation spectra, which may provide evidence of the reheating~\cite{Jaeckel:2020oet,Jaeckel:2021ert,Jaeckel:2021gah,Jaeckel:2022osh}.

\section*{Acknowledgement}
This work was supported by JSPS KAKENHI Grant Nos.  20H05851 (W.Y.), 21K20364 (W.Y.), 22K14029 (W.Y.), and 22H01215 (W.Y.).

\bibliography{ref}

\end{document}